# Concurrent Changes to Hadley Circulation and the Meridional Distribution of Tropical Cyclones

Joshua Studholme[1,*] and Sergey Gulev[1,2]

1. *Department of Physics, Shirshov Institute of Oceanography, Russian Academy of Sciences*

2. *Moscow State University*

*\* Corresponding author address: Joshua Studholme, Shirshov Institute of Oceanology, RAS.
36 Nakhimovsky ave. 117997. Moscow. Russian Federation.
Email: josh.studholme@gmail.com*

ABSTRACT

Poleward trends in seasonal-mean latitudes of tropical cyclones (TCs) have been identified in direct observations from 1980 to present. Paleoclimate reconstructions also indicate poleward-equatorward migrations over centennial to millennial timescales. Hadley circulation (HC) is often both implicitly and explicitly invoked to provide dynamical linkages to these shifts, although no direct analysis of concurrent changes in the recent period has been presented. Here the observational TC record (1981-2016) and the ERA-Interim, JRA55 and MERRA2 reanalyses are studied to examine potential relationships between the two. A zonally-asymmetric HC is defined by employing Helmholtz theory for vector decomposition and this permits the derivation of novel HC diagnostics local to TC basins.

Coherent variations in both long-term linear trends and detrended interannual variability are found. TC genesis and lifetime maximum intensity latitudes share trend sign and magnitude with shifts in local HC extent, with rates being ~0.25±0.1 °latitude decade$^{-1}$. Both these lifecycle stages in hemispheric means and all Pacific TC basins, as well as poleward-extreme North Atlantic lysis latitudes, shared ~35% of their interannual variability with HC extent. Local HC intensity is linked only to eastern North Pacific TC latitudes, where strong local overturning corresponds to equatorward TC shifts. Examination of potential dynamical linkages implicates La Niña-like sea surface temperature gradients to poleward HC termini. This corresponds to increased tropical and reduced subtropical vertical wind shear everywhere except in the North Atlantic and western North Pacific, where the opposite is true. These results quantify a long-hypothesized link between TCs and the large-scale oceanic-atmospheric state.

# 1. Introduction

Tropical cyclones (TCs) play an important role in the climate system and are useful indicators of climate variability. Since there are strong regional signals of ongoing climate variability in both the Atlantic and Indo-Pacific Oceans (e.g. Rhein et al. 2013; Li et al. 2015), the responses of TCs to these signals are of great interest. TCs are also hypothesized to play an active role in influencing climate at both the local and global scales (e.g. Emanuel 2008; Manucharyan et al. 2011; Fedorov et al. 2010, 2013; Huang et al. 2017). TCs represent important extreme climate events and impact financial markets, especially the insurance and reinsurance industries (Pielke 2007; Reed et al. 2015). In this respect, TC tracks are of great physical and societal relevance.

Numerous approaches have been used to quantify TC tracks and their variability in climate records and numerical simulations. A binary classification of 'straight-moving' (typically zonal propagation) versus 'recurving' (moving poleward after genesis and curving back upon themselves) is very widely employed. Other metrics include point and track densities (e.g., Elsner et al. 2012), trajectory mass moments (e.g., Nakamura et al. 2009), and probabilistic clustering (e.g., Gaffney et al. 2007; Kossin et al. 2010). A description of TCs using zonal and meridional averages along the lifecycle has been employed in a number of recent analyses (e.g., Kossin et al 2014, 2016; Wu et al. 2015; Wang et al. 2016). This approach has particular merit when relating TC activity to hemispheric climate modulations.

Considerable intraseasonal and interannual variability as well as long-term trend in TC tracks is well documented, although substantial issues remain for analysis. Difficulty persists since available timeseries are short and changes in operational centers and practices occur throughout the available record (e.g., Sandgathe 1987; Chia and Ropelewski 2002; Landsea 2015; Klotzbach and Landsea 2015; Landsea et al 2010). A poleward migration in the seasonal-mean latitude of TC lifetime-maximum intensity (LMI) in both hemispheres has been identified with rates of 0.48 and 0.56 °latitude decade$^{-1}$ between 1982 and 2012, in the Northern and Southern Hemisphere (NH, SH) respectively (Kossin et al. 2014). Daloz and Camargo (2017) found LMI migration corresponds to poleward migration of Pacific TC genesis. Changes in TC latitudes, linked to recurvature, have also been identified regionally in both the western North Pacific and North Atlantic (Wu and Wang 2004; Kossin et al. 2010; Ha et al. 2014; Daloz et al. 2015; Mei and Xie 2016; Hart et al. 2016; Kossin et al. 2016). Over centennial and millennial timescales, TC latitudes have been shown to migrate poleward and equatorward at various times in numerous paleoclimatic reconstructions (e.g. Baldini et al. 2016; Hengstum et al. 2016).

Coincident with these poleward migrations in seasonal-mean TC latitudes, there has been an expansion of the dominant feature of the large-scale tropical atmospheric flow, the Hadley Circulation (HC, e.g. Lucas et al. 2014). Reported magnitudes for the migration of the HC poleward extent are approximately 0.5 to 0.81 °latitude decade$^{-1}$ (e.g. Hu and Fu 2007; Stachnick and Schumacher 2011; Davis and Rosenlof 2012; Allen et al. 2012; Nguyen et al. 2013; Davis and Birner 2013). There is thus some amount of consistency between the reported poleward trends in TC latitudes and HC extent.

TC tracks are principally determined by the large-scale mean tropospheric flow and to a lesser extent, genesis locations (e.g., Riehl and Shafer 1944; Colbert and Soden 2012; Colbert et al. 2015). Genesis locations, i.e. positions where disturbances reach TC intensity, are chiefly determined by sea surface temperature (SST, e.g. Emanuel 2005; Vecchi and Soden 2007b; Elsner et al. 2008; Vecchi et al. 2008; Villarini et al. 2010; Ramsay and Sobel 2011; Villarini et al. 2012; Defforge and Merlis 2017) and vertical wind shear (VWS) (e.g., Emanuel and Nolan 2004; Vecchi and Soden 2007a; Vimont and Kossin 2007; Kossin and Vimont 2007; Kossin 2017). An influence of second order importance is TC recurvature, which is an intrinsic property resulting from a differential horizontal advection of planetary vorticity, sometimes referred to as beta drift (Bin et al. 1998; Chan and Chan 2016). Since TC tracks are influenced by their environmental large-scale winds and SST, links to large-scale structure of the tropical circulation are of great interest (Latif et al. 2006; Kossin et al. 2010; Vecchi et al. 2013a).

The HC is often implicitly and explicitly invoked to provide a mechanistic interpretation of the poleward-equatorward migration of seasonal-mean TC latitudes. Kossin et al. (2014, 2016) relate the recent poleward migration of TCs to increases in seasonal-mean VWS in the tropics and decreases in the subtropics by subtracting composites of an early and late period, thus implicitly invoking HC as it is known to have migrated poleward over the period. This pattern corresponds to an increase in TC potential intensity (connected to tropical SSTs and ocean-atmosphere thermal disequilibria). They went on to indeed hypothesize that these changes are directly linked to the well-documented expansion of the HC (e.g. Fu et al. 2006; Lucas et al. 2014).

Furthermore, Baldini et al. (2016) hypothesized that the consistent poleward migration of North Atlantic TCs that they found dating back to 1550 A.D. in stalagmite-derived paleoclimate records is linked to concurrent HC changes. They propose a mechanistic link of the HC's impact of TC latitudes through displacement of the Bermuda High. On even longer timescales, Hengstum et al. (2016) used a 3,000-year sedimentary record to conclude that the HC has likely contributed to the meridional distribution of intense TCs in the North Atlantic over millennial to centennial-scales.

Despite comparable magnitudes in poleward trends and the conceptual dynamical linkages, an explicit analysis between TC latitudes and the HC in the direct observational record has not yet been performed. Recently, local Hadley cells have been diagnosed from the divergent component of the meridional winds (e.g. Zhang and Wang 2013; Schwendike et al. 2014; Schwendike et al. 2015; Zhang and Wang 2015; Nguyen et al. 2017). This method is particularly appropriate for studying TCs in relation to the HC as HC is notably zonally asymmetric. Zhang and Wang (2013, 2015) used this diagnostic to examine how the Atlantic and eastern North Pacific local HC effects TC frequency and intensity.

Here we present a novel analysis of TC-HC concurrencies using the 35-year observational TC record and three reanalysis products to address the hypothesized connection directly. Specifically, the analysis has the following aims:

1. Identify poleward-equatorward migrations in seasonal-mean TC latitudes along the cyclone lifecycle (genesis, LMI and lysis).

2. Diagnose the local HC, and compute and discuss the magnitudes of long-term linear trends in these diagnostics directly in relation to concurrent trends in seasonal-mean TC latitudes.

3. Directly relate interannual HC modulations to TC latitudes irrespective of any long-term linear trends in either.

As far as the authors are aware, this is the first attempt to make such an explicit covariance analysis between TC and local HC. The structure of this paper is as follows: in section 2 the data and methodology are detailed. Section 3 describes changes in TC meridional distribution over recent decades. Sections 4 and 5 analyze the HC meridional extent and intensity and their relationship to TC latitudes respectively. A brief exploration of potential dynamical linkages is provided in section 6 before a summary in section 7.

**2. Data and Methodology**

*a. Observational TC data*

TC track data is taken from the International Best Track Archive for Climate Stewardship (IBTrACS) v03r10 (Knapp et al. 2010). This archive is maintained by the United States' National Oceanic and Atmospheric Administration (NOAA) National Climatic Data Center and is considered the most complete set of global historical TC data (e.g., Kossin et al. 2014; Walsh et al. 2016; Kossin 2017).

Uncertainties in TC observational records have a strong impact on analysis reliability (e.g. Landsea et al. 2010; Kossin et. al. 2014; Landsea 2015; Klotzbach and Landsea 2015; Daloz et al. 2015). Data have improved since 1980 with global satellite coverage but even post-satellite records contain significant uncertainties (Ren et al. 2011; Barcikowska et al. 2012; Torn and Snyder 2012; Landsea and Franklin 2013). The length of reliable records is important for trend detection since natural variability at decadal and longer scales is high (Sobel et al. 2016). Inhomogeneities also hamper analyses, and there is particular observational bias towards times and locations when TCs are within a forecaster's 'realm of interest' (i.e. the system is intense and there is landfall risk). This bias chiefly affects records of genesis and lysis as this is when cyclones are typically weak and far from landfall (Kossin et al. 2014). To maximize robustness and analyze sensitivity, where possible it is optimal to use more than one dataset (Camargo and Sobel 2010; Schreck et al. 2014).

We use IBrACS-WMO (the archive's own homogenization product) and another aggregation combining National Hurricane Center data (aka. RSMC Miami, aka. HURDAT2), which covers the Atlantic and eastern North and Central Pacific, and the JTWC, which covers all other basins (together referred to as NHC-JTWC). We employ both IBTrACS-WMO and NHC-JTWC, in all basins except the North Atlantic and eastern North Pacific (where the NHC *is* the WMO observational agency). We only show analysis using the IBTrACS-WMO data for the North Atlantic and eastern North Pacific. We note that using two datasets does not completely exclude bias as both are depend upon their own set of potentially flawed subjective operational procedures. Different agencies use different wind averaging periods. Following Schrech et al. (2014), we use 1-min winds and thus convert where appropriate after their methodology.

NH and SH means are constructed from both IBTrACS-WMO and NHC-JTWC. We deem this a valid approach in the NH since the majority of TCs there occur in the western North Pacific where two independent records exist. Regional boundaries used in this study (defined following WMO guidelines) and all TCs (by key lifecycle point) are shown in figure 1a.

For consistency and reproducibility, we examine periods when tracks are considered either 'tropical' or 'subtropical' as indicated by archive flags, a categorization that is standardized by IBTrACS. We also use only the 'main' (archive's terminology) time series of all TCs as denoted by archive flags. The inclusion of wind speed data is variable across ocean basins and time. Again for consistency, we define genesis with a constant wind speed threshold. Some tracks in the archive do not include any wind speed data and are thus excluded from the analysis.

We consider three specific stages covering the TC lifecycle: 'genesis', 'lifetime maximum intensity' (LMI), and 'lysis' (figure 1b,c). Here, the genesis location is defined as the location of the TC center when it first reaches tropical storm intensity (defined as maximum sustained wind speed $\geq$ 18 m/s, 33 kts). If a TC's first recorded wind speed is above this threshold, this first point is taken as genesis. Such tracks are <1% of the archive and their inclusion/exclusion does not affect the conclusions reached. LMI is the time at which the highest magnitude maximum sustained wind speed over the entire lifecycle is recorded. LMI can be viewed as the most robust measure as this depends solely on the maximum intensity which will occur well into the lifecycle whereas genesis and lysis will be more sensitive to operational changes and forecaster subjectivity. Lysis location is the position of the track when its 'nature' (archive's terminology) is last recorded as tropical. In practice, TCs at this point in their lifecycle may indeed experience actual lysis, i.e. a disintegration of the defining coherent warm core, or may undergo extratropical transition (ET). Determining ET in observational data is non-trivial (e.g., Studholme et al. 2015); thus, aggregating actual warm core disintegration and ET into the category of 'lysis' in this analysis is considered to be a reasonable and meaningful approximation.

*b. Hadley circulation extent and intensity diagnostics*

Estimates for HC extent and intensity are derived from three reanalyzes: ERA-

Interim (Dee et al. 2011), JRA55 (Kobayashi et al. 2015) and MERRA2 (Gelaro et al. 2017) using three-dimensional monthly means of daily means on their native grids (3/4° x 3/4° for ERA-Interim, 1/2° x 1/2° for JRA55, 1/2 x 2/3 degrees for MERRA2) at all available pressure levels. It is important to use multiple reanalyzes as considerable variation in HC has been shown to exist between them (e.g. Nyugen et al. 2013). These data are then linearly interpolated onto 1/4° horizontal resolution and 10 hPa vertical resolution in pressure coordinates to aid comparison.

To diagnose the local HC, we employ the Helmholtz decomposition following Schwendike et al. (2014). This takes advantage of the fact that a vector field in $\mathbf{R}^3$ may be unambiguously partitioned into two contributions: (i) a divergent (aka. irrotational) component where $\nabla \times \mathbf{V} \equiv 0$ and (ii) a non-divergent component where $\nabla \cdot \mathbf{V} \equiv 0$ (e.g. Bourne and Kendall 1992). This is necessary for defining the zonally asymmetric HC since the commonly-used Stokes streamfunction computed using hemispheric zonal-mean meridional velocities, $[v]_0^{2\pi}$, depends upon the assumption of nondivergence, i.e. $\partial v/\partial y + \partial \omega/\partial p = 0$ (using standard geophysical notation; e.g. Oort and Rasmusson 1970; Oort and Yienger 1996). It therefore requires there to be no zonal component to the net mass flux. In a zonal-mean of the atmospheric wind field over an entire hemisphere this condition is trivially satisfied. However, between any two arbitrary zonal bands $\lambda_1$ and $\lambda_2$, where $\lambda_2 - \lambda_1 < 2\pi$, this ceases to be true. Over such regional domains, the net zonal mass flux is non-negligible and the streamfunction may not be computed from $v$. Using the divergent component of the Helmholtz-decomposed winds ensures that this condition is satisfied. In the interest of brevity, we direct the reader to Schwendike et al. (2014) for a full derivation of this approach. Summarizing, vertical motion can thus be written as:

$$\omega = \omega_\phi + \omega_\lambda, \quad [1]$$

where $\omega$ is vertical motion in pressure coordinates and subscripts $\phi$ and $\lambda$ denote the vertical velocity in the meridional and zonal planes respectively. The component $\omega_\phi$ can be viewed as the meridional overturning circulation (i.e. Hadley circulation) and the component $\omega_\lambda$ as zonal overturning, (Walker circulation). In this analysis, we are only interested in $\omega_\phi$ which satisfies:

$$\frac{\partial v_\chi}{\partial y} + \frac{\partial \omega_\phi}{\partial p} = 0, \quad [2]$$

where $v_\chi$ is the divergent component of the meridional velocity. Although this method assures $\omega_\phi$ and $\omega_\lambda$ are two unique orthogonal circulations, it does not imply independence between the zonal and meridional components of the flow, and such a coupling has been implicated in interbasin associations (e.g. Chiang et al. 2000; Chiang et al. 2002). The meridional streamfunction in spherical coordinates for the local HC can thus be derived as (Zhang and Wang 2013; Nyugen et al. 2017):

$$\psi(\phi, p) = \frac{2\pi a}{g} \int_0^p [v_\chi]_{\lambda_1}^{\lambda_2} \cos\phi \, dp, \quad [3]$$

where *a* is the mean radius of the planet and *g* is gravitational acceleration.

The integration is taken from the top of the atmosphere which yields positive

(negative) values for annual-mean $\psi$ in the NH (SH) tropics. This corresponds to a clockwise (counter-clockwise) circulation when viewed from the East. We compute $\psi$ between zonal bands corresponding to different TC basins: North Atlantic (20°W-70°W), eastern North Pacific (105°W-150°W), western North Pacific and South Pacific (130°E-180°E), and South Indian (50°E-110°E). We verified all results against reasonable perturbations (i.e. ±10°) around these boundaries and they had no meaningful effect on results.

The HC is characterized with two diagnostics derived from $\psi$: HC extent and intensity. Traditionally HC extent has been defined as the $\psi=0$ isoline (e.g. Oort and Yienger 1996). However, in the zonally asymmetric HC this line is not necessarily crossed (figure 2). Nguyen et al. (2017) first encountered this issue when looking at regional HC in the SH. They devised a new definition of the edge of the local HC, defining it as where the overturning weakens to a specified threshold percentage of the maximum overturning within the tropical cell. The edge of the HC is therefore taken as the average position of the weakened-peak-value between 700-400 hPa. We use 800-400 hPa as we found that shallow cells, particularly in the North Atlantic, were better detected by including lower levels without losing skill elsewhere. Nguyen et al. (2017) conducted sensitivity tests with thresholds of 10, 15, 20, and 25%. We conducted our own tests and reach similar conclusions. Specifically, that while the absolute latitude of the HC edge is sensitive to the threshold value, the variability is not. Higher thresholds more robustly detect the HC edge and so here, as in Nguyen et al. (2017), we use 25%. We tested this definition against the traditional zero isoline for the hemispheric zonal-mean and it indeed captures the same variability (correlation coefficient 0.95). More details of the HC diagnostic algorithm are in Appendix A.

HC intensity is taken as the vertically averaged maximum value of $\psi$ between 900 and 200 hPa in each overturning cell. The interested reader is directed to Nguyen et al. (2013) for an analysis of global zonal-mean HC in reanalyses and Nguyen et al. (2017) for a discussion of local HC variability in the Southern Hemisphere.

Timeseries for local HC diagnostics are produced by applying the diagnostic algorithm to $\psi$ computed with monthly-mean $[v_\chi]_{\lambda_1}^{\lambda_2}$. HC has a strong seasonal cycle, distinct not only between hemispheres but also ocean basins. For this reason, when taking seasonal means of HC diagnostics, we weight each month's value of HC extent and intensity by the corresponding TC count in that basin or hemisphere for that specific month.

*c. Study period*

The analysis covers 1981 to 2016. 1981 is the start of global coverage for both TC datasets and approximates the start of satellite data coverage. Note that the IBTrACS-WMO record of the northern Indian Ocean begins in 1990. As a result, and as meteorological conditions that affect TCs in this ocean are markedly different from elsewhere, we exclude North Indian TCs. 2016 is the last complete year currently included in the TC archive. We define JASO and JFM as TC seasons in the NH and SH respectively, corresponding to months with ~75% of all TC activity there. None of the time series in this analysis exhibited autocorrelation as determined by the Durbin-Watson test statistic.

**Results: Changes in meridional TC distribution in recent decades**

*a. Reference climatology of TCs*

Figures 1 b and c show climatological distributions of all TC lifecycle points in the WMO and NHC-JTWC datasets respectively. TC genesis is generally confined between the Equator and ±20°N and is more tightly bound in the SH. This is possibly due to differences in operational procedures (e.g. Hodges et al. 2017). The SH TC lifecycle follows a strict meridional progression that is zonally symmetric across the South Indian and South Pacific oceans. SH TCs reach LMI between 10 and 20°S and experience lysis poleward of this. WMO and NHC-JTWC notably differ in detecting lysis latitudes in the South Pacific. The WMO archive tracks cyclones for much longer, as south as 60°S, while NHC-JTWC terminates tracks around 40°S. In all other respects, the two datasets are very consistent with one another in terms of spatial distributions.

In the NH, the TC lifecycle is notably zonally asymmetric. TC genesis in the western North Pacific occurs very close to the Equator, while in the eastern North Pacific and North Atlantic it is bounded on the equatorward side at 10°N. This might reflect differences in source disturbances leading to genesis such as Easterly waves. Eastern North Pacific TC genesis does not occur more poleward than approximately 15°N, while in the western North Pacific and North Atlantic, it can occur as far northward as 40°N (again this strongly dependent on observational procedures). LMI latitudes shows a similar spread between 10 and 40°N in both the western North Pacific and North Atlantic but are not observed poleward of 20°N in the eastern North Pacific. NH lysis latitudes are very zonally asymmetric. In the western North Pacific, they are bimodal with peaks at approximately 20 and 40 °N. In the eastern North Pacific, TC lysis occurs at approximately 25°N ± 5°, with a tendency to occur farther north as cyclones propagate eastward. In the North Atlantic, lysis is distributed across the entire ocean but most prominently occurs along the North American East Coast soon after landfall or at approximately 40°N.

Long-term regional statistics are given in table 1. The NH exhibits well over twice the number of TCs than the SH. According to WMO estimates, the NH experiences an average of 59 TCs per year versus 25 in the SH. This is partly related to differences in inclusion of sub-tropical cyclones (e.g., Hodges et al. 2017). However, it predominately results from the presence of an additional active TC basin (three in the NH vs. two in the SH) and also because the western North Pacific is a particularly active, with 25 TC on average per year according to WMO estimates.

The western North Pacific thus largely dominates the NH average. This is most probably due to the large area of very warm SSTs there. Even for seasonal statistics, differences in TC records between datasets (WMO vs. NHC-JTWC) are apparent.

*b. Meridional quantile regressions*

To examine changes in mean meridional TC locations, we compute quantile regressions (e.g. Elsner et al. 2008) in both hemispheres and all basins for TC genesis, LMI, and lysis. This is done by fitting linear ordinary least squares models to timeseries of TC latitudes for percentiles across the meridional distribution. These are characterized by a regression coefficient, interpreted as linear poleward trend, and the corresponding 95% confidence limits. We use the model implementation of the Python Statsmodels environment (Seabold and Perktold 2010).

Figure 3 shows results for hemispheric domains. We note that all TC instances are equally weighted regardless of ocean basin (i.e. all TCs are uniformly weighted) and again that, in the North Atlantic and eastern North Pacific, WMO and NHC-JTWC data are the same. Signs of LMI trends (figures 3b, c) are in agreement with Kossin et al. (2014). We find a poleward migration of the annual mean TC LMI latitudes for both hemispheres. In the SH, this trend is robust at the 95% significance level in both datasets. In the NH, it is only robust at this confidence level in WMO data and even so, is very slight. Kossin et al. (2014), from an alternative dataset aggregation and time period (1982-2009, IBTrACS, selecting sources for each TC with the most poleward LMI), reported tendencies in LMI of 53 and 62 km per decade in the NH and SH, respectively. We find trends of 0.1 and 0.45 °latitude decade$^{-1}$, which correspond to 11 km and 50 km per decade, respectively, between 1981 and 2016. Given this comparison to Kossin et al.'s 1982-2009 trend estimates, it would stand to reason that while the SH trend has continued on, the NH LMI poleward migration has halted in the most recent period. This is confirmed by inspection of timeseries (see figure 8). There is also a notable agreement between datasets in the hemispheric mean trend estimates for poleward migration of LMI in both hemispheres (shown as red horizontal lines in figure 3).

Both datasets also reveal a robust poleward migration of the most equatorward (below the 50$^{th}$ percentile) NH LMI locations (significant at the 95% level in the WMO). By contrast, there are considerable uncertainties concerning migration at the most poleward (above the 50$^{th}$ percentile) TC LMI latitudes in the NH. In SH LMI, there is no robustness in trends below the 40$^{th}$ percentile (most equatorward LMI locations), but both datasets at the 95% confidence level show that the 60% most poleward LMI latitudes are migrating further southward. Again, this phenomenon occurs at a rate of 0.4°latitude decade$^{-1}$.

NH TC genesis is migrating poleward at a faster rate than the LMI, with a rate of 0.4°latitude decade$^{-1}$. This is confirmed by the two datasets and is robust at the 95% significance level. This trend is dominated by the poleward migration of the most equatorward (below the 50$^{th}$ percentile) half of genesis latitudes, while the trend for the poleward half is not statistically robust. In the SH, while the NHC-JTWC dataset shows a robust trend in the mean genesis latitude of a poleward migration at 0.43°latitude decade$^{-1}$, this trend is not identified in the WMO data.

Quantile regressions for NH ocean basins (Figure 4) show robust poleward migration in seasonal-mean LMI in WMO data of 0.25°latitude decade$^{-1}$ in the western North Pacific, but no such trend in NHC-JTWC. We also find a robust poleward migration of the same magnitude in the lower 50$^{th}$ percentile of seasonal genesis latitudes in both data sources. At the same time, we find no robust trends in the North Atlantic, and this is consistent with Wang et al. (2016) and Kossin et al. (2014) results. In the eastern North Pacific, the poleward migration of TC genesis at 0.45 °latitude decade$^{-1}$ is robustly consistent across the distribution. This trend indicates a wholesale poleward shift in TC genesis in this region and compliments with the genesis shift in the western North Pacific. No such corresponding trend is found in LMI. We find that the core of the eastern North Pacific lysis meridional distribution (20$^{th}$ – 70$^{th}$ percentiles) follows a statistically significant equatorward trend of 0.33°latitude decade$^{-1}$.

In the South Pacific, we find a statistically robust mean LMI poleward migration in both datasets of 0.42°latitude decade$^{-1}$ (figure 5a). We also find a poleward migration in mean genesis latitudes in NHC-JTWC of 0.58 °latitude decade$^{-1}$ but it is not significant in the WMO data. These trends are matched in the South Indian (figure 5d, e, f). We find statistically significant trends in seasonal mean LMI in both datasets with differing magnitude: WMO has 0.5 °latitude decade$^{-1}$ and NHC-JTWC 0.45°latitude decade$^{-1}$.

Taken in aggregate, these trends can be summarized by four major points:

- A poleward migration of seasonal-mean TC LMI occurs in both the NH (0.1°latitude decade$^{-1}$) and SH (0.45°latitude decade$^{-1}$) in unweighted hemispheric averages.

- A Pacific-wide poleward migration of seasonal-mean TC genesis occurs (approximately 0.45°latitude decade$^{-1}$).

- An equatorward shift in seasonal-mean eastern North Pacific lysis by 0.3°latitude decade$^{-1}$.

- Notable uncertainties remain despite efforts for dataset homogenization and ever-lengthening time series.

**4. Results: Hadley circulation extent and comparison to TC latitudes**

The climatological divergent winds and meridional overturning in the ERA-Interim reanalysis (figure 2) shows that HC extent varies notably from basin to basin. It is particularly more poleward in western Pacific. Three main centers of action are clear: over the African continent, the Indo-Pacific region and the Americas. The Intertropical Convergence Zone (ITCZ) is clearly seen with values of $\bar{\psi}$ at 500 hPa approaching zero and convergence in the near-surface wind vectors.

*a. Long-term Trends*

Persistent poleward shifts over the second half of the last century in HC extent are well documented and observable in a number of diagnostics, with the strongest trends identified in the 1980s and 1990s (see review by Lucas et al. 2014). Figure 6 shows the annual and TC-seasonal mean timeseries derived for this analysis (corresponding linear trends 1981-2016 in table 2). In the annual hemispheric means across the whole period, we find no significant linear trend in any of the three reanalyses in the NH (figure 6a). In the SH, the three reanalysis products disagree even over the sign of the trends (figure 6b). Trends computed between 1981 to 2005 show a clear poleward shift of ~0.8°latitude decade$^{-1}$ in the NH and ~0.5°latitude decade$^{-1}$ in the SH which agrees with estimates derived from streamfunction methods over a similar time period (e.g. Hu and Fu 2007; Johanson and Fu 2009; Allen et al. 2012). Between 1999 and 2009, the annual-mean HC extent in both hemispheres (figure 6a,b) is remarkably stable as noted in other studies (e.g. Stachnik and Schumacher 2011; Davis and Rosenlof 2012; Nguyen et al. 2013; Davis and Birner, 2013). The period 2010 to 2016 shows a strong equatorward tendency in the NH, this being the counterbalance to the earlier poleward shifts that overall results in the absence of linear trend between 1981 – 2016 as a whole.

Trends for time series averaged over TC seasons and weighted for TC counts

(figure 6 c,d) show poleward shifts in both hemispheres of 0.3°latitude decade$^{-1}$. These trends are very robust in the SH with all reanalyses agreeing and trends exceed 95% confidence. Interannual variability is much greater in the NH, this being driven by the western North Pacific local HC (figure 6 g). In individual TC basins, there are no statistically robust poleward shifts consistent in all reanalyses (figure 6 e,f,g,h; table 2).

Comparing HC and TC poleward trends reveals reasonable overall coherence within a shared relatively tight range of estimated magnitudes. Summarising, in all but a few cases (in particular in the Northern Pacific), HC and TCs have tended poleward together with rates ~0.25 ±0.1°latitude decade$^{-1}$ (table 2, figures 3,4,5). TCs in the SH exhibit the strongest poleward migration of 0.45°latitude decade$^{-1}$ at LMI while over the same period SH HC has shifted poleward at a mean rate of 0.3°latitude decade$^{-1}$. In the NH, while the HC has also shifted poleward at 0.3°latitude decade$^{-1}$ (albeit less conclusive and with much higher interannual variability), TC LMI has shifted at a mean rate of approximately 0.1°latitude decade$^{-1}$.

### b. *Shared Interannual Variability*

When we directly regress detrended seasonal-mean TC latitudes onto seasonal-mean local HC extent, we find robust statistical covariance across the TC lifecycle occurring irrespective of long-term shifts (Figure 7). In general, we find Pacific and North Atlantic TC genesis shares variability with HC extent (r=0.6, figure 7a). The covariance strength is lower for LMI (r=0.5, figure 7b) and negligible for lysis (r < 0.4, figure 7c). Different reanalysis products generally produce similar results, but there are some differences between the two TC observational records.

In particular, the covariance between genesis and LMI latitudes and HC extent appears to be predominantly a Pacific phenomenon. At genesis, although notable covariance is found with North Atlantic TC latitudes, $R^2$ values are half that of the NH Pacific basins (figure 7a). In the South Indian, there is no notable relationship whatsoever.

Two hemispheric-mean timeseries demonstrate the covariance strength itself has varied over the period (figure 8). For NH HC extent and TC genesis latitudes (Figure 8a) the period 1981-1991 shows remarkable common interannual variability (rolling correlations of 0.9). During the 1990s the correlation drops to 0.5 as the HC shifts poleward nearly year-on-year and genesis latitudes remain remarkably consistent. From 2000 onwards, the high correspondence returns. A similar situation is demonstrated for SH TC LMI latitudes (figure 8b).

This set of regressions is repeated for the equatorward and poleward parts of the seasonal TC latitudes but are not produced here since they are more or less consistent with the means. The exception is that the most poleward quartile [ P(75) ] of North Atlantic lysis latitudes are found to significantly covary with local HC extent (r=0.62). As in the hemispheric case, we see that the observed covariance is time dependent (figure 9). In particular, it is very weak in the 1990s but strong in the 1980s and 2000s.

In aggregate, these results indicate that TC genesis and LMI latitudes share long-term trend sign and magnitude with concurrent shifts in local HC extent with rates around ~0.25 °latitude decade$^{-1}$. Seasonal-mean TC genesis and LMI mean latitudes in both hemispheres and in all Pacific basins, as well as poleward-extreme North Atlantic lysis latitudes, share ~35% share of their interannual variability with HC extent.

## 5. Results: Hadley Circulation Intensity and Comparison to TC Latitudes

### a. *Long-term Trends*

HC intensity has been subject to notable debate in the recent past. Thermodynamic scaling arguments predict a weakening of convective mass flux in the tropics resulting from changes in atmospheric humidity (Held and Soden 2006). This is consistent with long-term [ *O(100 yrs)* ] model simulations, although such a weakening has been found to occur preferentially in the zonal (i.e. Walker) circulation rather than in the HC (Vecchi and Soden 2007c). Analyses of HC intensity over the past thirty years have found weak signals of either increasing, decreasing, or no change in estimates and large uncertainties exist depending on datasets, methodology and time periods (e.g. Lau and Kim 2015).

Nguyen et al. (2013) studied hemispheric-mean HC in eight reanalyzes, including ERA-Interim but not JRA55 and MERRA2, for 1980-2009 using streamfunction methods. They found intensification or weakening in NH HC intensity ranging from -0.5 to 12.7 x $10^9$ kg s$^{-1}$ decade$^{-1}$. In the SH, trends ranged from -3.2 to 9.0 x$10^9$ kg s$^{-1}$ decade$^{-1}$. Between 1981 and 2009, annual-means trends for the present analysis (figure 10a, b) in the NH range from -0.1 to 6.8 x$10^9$ kg s$^{-1}$ decade$^{-1}$ and in the SH from 0.4 to 5.3 x$10^9$ kg s$^{-1}$ decade$^{-1}$. Thus, our estimates largely agree with Nguyen et al. (2013) despite different HC algorithms and reanalysis products.

All HC intensity timeseries exhibit very considerable interannual variability. Annual-mean NH HC shows persistent intensification (figure 10a, table 3) while the seasonal-mean (figure 10 c) shows trend ambiguity until 2005 when ERA-Interim and JRA55 indicate intensification and MERRA2 shows rapid weakening. This divergence between reanalyses seems to arise in the western North Pacific (figure 10 g). Over the entire period, all reanalyzes show slight linear strengthening in the SH (figure 10d, table 3), while it is not evident in individual TC basins. Thus, the weakening could be attributed to regions that do not experience TCs such as the African continent or eastern South Pacific. The seasonal-mean NH HC appears to have weakened with a trend of approximately -10 x$10^9$ kg s$^{-1}$ decade$^{-1}$ over the last five years in all three reanalyses.

*b. Shared Interannual Variability*

Regressions between detrended seasonal-mean HC intensity and seasonal-mean TC latitudes (figure 11) do not indicate any robust covariance across hemispheric means in all but one ocean basin. We find a robust link between seasonal-mean eastern North Pacific HC intensity and seasonal-mean latitudes of TC genesis and LMI in all reanalysis products. The same conclusions also hold for the most equatorward and poleward edges of genesis and LMI latitudes.

Corresponding time series of TC genesis equatorward extreme latitudes (Figure 12) shows strong, time-invariant negative covariance (r= -0.75). A more intense eastern North Pacific HC is associated with equatorward shifts in TC genesis and LMI there. The flow in this basin is strongly influenced by the Walker circulation, orthogonal to the HC, which makes it unlike other basins. This zonal influence might perhaps be related to this anomalous negative covariance. Note also that over this period there is no trend consistent across reanalysis products for an intensification or weakening of the local overturning circulation (table 3) but there is a persistent and very robust, wholesale poleward shift in the latitudes of eastern North Pacific tropical cyclogenesis.

Summarising, long-term HC intensity trends (table 3) show that although there is broad consensus that over the period HC has generally intensified while TCs have migrated poleward, the spread in HC intensity trend estimates is so large ($>12 \times 10^9$ kg s$^{-1}$ decade$^{-1}$) that concluding any coherent relationship is somewhat dubious.

Results indicate there is no common interannual variability between seasonal-mean TC latitudes and HC intensity, except in the eastern North Pacific.

6. **Potential dynamical linkages**

Significant uncertainties remain in fundamental understanding of both TCs' relationship to large-scale climate dynamics, and the HC despite it being the oldest known large-scale atmospheric circulation, (e.g. Gastineau et al. 2011; Levine and Schneider 2011; Holton et al. 2013; Gleixner et al. 2013; Zhan et al. 2017; Yan et al. 2017). It is well understood that TCs are strongly sensitive to both SSTs and VWS (e.g. Murakami et al. 2011; Kossin et al. 2014; Kossin et al. 2016; Yan et al. 2017). Regressing the hemispheric-mean HC extent diagnostic against seasonal means of these quantities from ERA-Interim reveals strongly coherent spatial patterns implicated in potential underlying dynamical linkages (figure 13).

We see a more poleward HC is linked to reduced meridional SST gradients in all ocean basins and warm SST anomalies at TC latitudes in the North Atlantic, western North Pacific and South Pacific. There are even slight warm anomalies in the eastern North Pacific and South Indian SSTs around TC locations despite generally being cooler in these basins on the whole (figure 13a). This pattern is remarkably reminiscent of a strong La Niña event (e.g. Lim et al. 2016) and links between an expanded HC and La Niña have been noted before (e.g. Seager et al. 2005; Seager et al. 2010). In idealized numerical experiments, HC extent has been shown to have a complex relationship to absolute SSTs and meridional SST gradients (e.g. Walker and Schneider 2006; Levine and Schneider 2011; Gastineau et al. 2011; Seo et al. 2014).

Tropical VWS generally increases and subtropical VWS generally decreases with a more poleward HC (figure 13b). Kossin et al. (2014) arrived at the same result by subtracting zonally averaged composites over 1980-1994 and 1995-2010, implicitly linking TC poleward migrations to HC. Here, we arrive at the same general conclusion but additionally reveal strong zonal asymmetries. In particular we see tropical (subtropical) VWS reduction (increase) in the North Atlantic and western North Pacific, i.e. the inverse of the zonally integrated pattern.

In the eastern North Pacific, we find more intense local HC strongly reduces equatorward VWS (figure 14b) and is linked to basin-wide warmer SST (figure 14a). This environmental change is linked to an equatorward shift in seasonal-mean TC latitudes. With this method, we are able to identify basin specific heterogeneity which may explain the differences we observe in trends and covariance. These linkages beg further analysis.

Previous works have identified large-scale dynamic and thermodynamic changes modulating TC meridional distribution (e.g. Kimberlain and Elsner 1998, Saunders and Lea 2008; Li et al. 2010; Kossin et al. 2010, Kossin et al. 2014; Gleixner et al. 2014; Kossin 2017; Yan et al. 2017). The El Niño Southern Oscillation (ENSO) in particular has been linked to changes in both TC distributions and HC (e.g. Wang and Chan 2002; Chia and Ropelewski 2002; Tao et al. 2012; Oort and Yienger 1996; Nyugen et al. 2013; Seager et al. 2005; Seager et al. 2010). In recent decades, SSTs have been trending towards a 'La Niña-like' anomaly pattern similar to the one identified in figure 13a (e.g. Lim et al. 2016).

## 7. Summary and discussion

We present an analysis comparing seasonal-mean poleward-equatorward migrations of TCs with concurrent changes in HC extent and intensity. Over the period 1981-2016 we find a poleward migration of TC LMI in the NH of $0.1°$latitude decade$^{-1}$ and in the SH of $0.45°$latitude decade$^{-1}$ in hemispheric averages. While the SH trend is comparable to the previously identified poleward migration there, the NH trend is about a quarter of the magnitude reported by Kossin et al. (2014) for the period 1982-2009. In the last five years, NH LMI latitudes have tended equatorward. We also observe a Pacific-wide poleward migration of seasonal-mean latitudes of tropical cyclogenesis (approx. $0.45$ °latitude decade$^{-1}$) and an equatorward shift in seasonal-mean eastern North Pacific lysis by $0.33°$latitude decade$^{-1}$. These results generally agree with estimates derived in other analyses (e.g., Kossin et al. 2014; Wang et al. 2016; Daloz and Camargo 2017).

Over the same period, both hemispheres' mean HC extent has shifted poleward at an approximate rate of $0.3°$latitude decade$^{-1}$. There are no robust poleward shifts in local HC extent that occur in all reanalyses. However, reanalysis products generally agree on the sign and magnitude of trends. Detecting local trends is challenging in part due to very high interannual variability. Over the entire period, HC intensity averaged over TC seasons has no consistent trend. Reanalyzes agree that hemispheric mean HC overturning has strengthened, although at the resolution of individual TC basins this is not evident. These findings are broadly in agreement with other works (e.g. Hu and Fu 2007; Johanson and Fu 2009; Stachnik and Schumacher 2011; Allen et al. 2012; David and Rosenlof 2012; Nguyen et al. 2013; David and Birner 2013).

After removing long-term trends, we find that latitudes of seasonal-mean TC genesis and LMI in the hemispheric mean and in the Pacific basins have approximately 35% of their interannual variability in common with HC extent. We also find that the poleward extremes in North Atlantic TC lysis latitudes and local HC extent share 40% of their detrended interannual variability. HC intensity has little-to-no relation to TC latitudes everywhere except in the eastern North Pacific where there is a strong negative covariance. As far as we are aware, this is the first attempt to derive statistics of this type. Thus, we find quantitative evidence for concurrent TC and HC meridional shifts at both interannual and long-term timescales. Notable issues remain with available data and uncertainty is considerable.

At the hemispheric scale, we find a more poleward HC is linked to reduced meridional SST gradients and warm SST anomalies at TC latitudes in such a way as to resemble a La Niña-like oceanic state. This corresponds to tropical (subtropical)

VWS generally increasing (decreasing). Locally, the inverse is observed in the North Atlantic and western North Pacific. Projections into the next century find trends towards weakened zonal SST gradients (i.e. El Niño-like) and in fact failed to produce the recently observed La Niña-like trends (e.g. Xie et al. 2010; Collins et al. 2013). Thus, it could well be that poleward-equatorward migrations in TCs and HCs reverse in the near future. However, there is significant model uncertainty in long-term ENSO simulations (e.g. Kociuba and Power 2015; Luo 2015; Rashid and Hirst 2016). Additionally, changes to TC latitudes relative to large-scale dynamic conditions such as HC may well change rates of extratropical transition (Mokhov et al. 2014).

Aside from outstanding dynamical questions, various issues surrounding diagnostics remain important. When studying uniformly-weighted hemispheric zonal-mean TC statistics as we do, inter-basin changes in TC frequency can impact results. It is known that such changes contribute equally to poleward migration as those of environmental changes (Kossin et al. 2014; Moon et al. 2015). Changes in genesis locations affect the number of recurving TCs and thus have a strong impact on lifecycle latitude metrics such as LMI (Wang and Chan 2002; Elsner and Liu 2003; Chan and Liu 2004; Camargo and Sobel 2005; Camargo et al. 2007; Yonekura and Hall 2014). Thus, while direct impact of large-scale SST patterns (e.g., ENSO) on TCs are mainly found in genesis locations, this signal is projected onto subsequent TC tracks and in the TC meridional distributions throughout the lifecycle (Yonekura and Hall 2014).

Some scholars have concluded that contemporary models are capable of simulating TCs with acceptable veracity (e.g. Zhao and Held 2010; Murakami et al. 2013, 2015; Han et al. 2016). However, using modeling to study HC, ITCZ dynamics and TC tracks remains a difficult undertaking since characteristic spatial scales are orders of magnitude apart. In this respect, there is much work to be done to improve physical understanding and projections of future HC and TC track changes (e.g., Levine and Scheider 2015; Daloz et al. 2015).

**Appendix A. Algorithms for Diagnosing Zonally-Asymmetric Hadley Circulation**

In a zonally-asymmetric HC, extent is notably harder to robustly identify than in the hemispheric zonal mean. In the latter, Hadley cell termini can be identified with the $\psi_{500\ hPa} = 0$ isoline whereas in the former, this line is not necessarily crossed at all. To account for this, we implemented and built upon the approach of Nyugen et al. (2017). We detail this in the methods section. In implementing this algorithm, it became clear that a few notable 'edge cases' exist where unexpected results are produced. We document these here for two reasons. First, it is clear that the specifics of any diagnostic algorithm have a strong effect upon the resultant time series. Therefore, explicitly demonstrating how they work is important for clarity. Second, we include this appendix to aid reproducibility.

The first case is when a second maximum in the overturning exists in the midlatitudes of the same sign as the tropical cell. The HC over the western Pacific warm pool region is a particular example of this (figure A.1a). This second higher-latitude maximum is often greater in magnitude than its tropical counterpart. The weakening threshold algorithm we implement here does a good job at identifying the terminus (blue vertical line at 28°N) despite this second maximum and the fact that the zero isoline is never crossed. This situation seems to occur when the local HC is merged or dominated by monsoonal flow in the boreal summer.

The second case we highlight is shown in figure A.1b. This occurs when the local Hadley cell is particularly weak and does not penetrate deep into the troposphere. Classical definitions of Hadley cell termini are defined relative to a vertical average between 600 and 400 hPa or a slice at 500 hPa (e.g. Stachnik and Schumacher 2011). Using an algorithm based on the weakening between 800 and 400 hPa captures these shallow weak overturning cells well (see blue NH vertical in figure A.1b).

As noted above, the absolute values for the latitude of HC extent that this algorithm detects are typically lower than the classical zero isoline variant. However, the variability of HC is better captured and configurations of atmospheric overturning that are beyond the classical definition's scope are accounted for.

*Acknowledgements.* The authors would like to thank Kevin Hodges very much indeed for the in-depth and helpful discussions and reading of the manuscript that have led to great improvements. Also, James Kossin and Carl Schreck III for very useful conversations and guidance on handling the many nuances of the tropical cyclone observational record and IBTrACS data. Likewise, we thank Hanh Nguyen very much for friendly discussions about diagnostic algorithms for Hadley circulation. We much appreciate the extremely helpful comments and suggestions of the anonymous reviewers and the editor John Chiang for having helped to improve the manuscript immensely. Finally, we thank NASA, ECMWF, and JMA for releasing their data to the public and the open-source Python and data-analysis community. This work was funded through the Agreement 14.W0331.006 with Ministry of Education and Science of the Russian Federation and grant no. 14-50-00095 from the Russian Science Foundation.**References**

Allen, R. J., S. C. Sherwood, J. R. Norris, and C. S. Zender, 2012: Recent Northern Hemisphere tropical expansion primarily driven by black carbon and tropospheric ozone. *Nature*, **485**, 350–354, doi:10.1038/nature11097. http://dx.doi.org/10.1038/nature11097.

Baldini, L. M., and Coauthors, 2016: Persistent northward North Atlantic tropical cyclone track migration over the past five centuries. *Sci. Rep.*, **6**, 37522, doi:10.1038/srep37522. http://www.nature.com/articles/srep37522.

Barcikowska, M., F. Feser, H. von Storch, M. Barcikowska, F. Feser, and H. von Storch, 2012: Usability of Best Track Data in Climate Statistics in the Western North Pacific. *Mon. Weather Rev.*, **140**, 2818–2830,35

# LIST OF TABLES

**Table 1.** Description of the IBTrACS TC count timeseries used in the analysis for both datasets: WMO | NHC-JTWC. Note that only one dataset is available in the North Atlantic and eastern North Pacific.

|  | N. Hemisphere | S. Hemisphere | W. North Pacific | N. Atlantic | E. North Pacific | S. Pacific | S. Indian |
|---|---|---|---|---|---|---|---|
| Total | 2079 \| 2202 | 885 \| 925 | 873 \| 1002 | 529 | 677 | 351 \| 358 | 534 \| 567 |
| Annual Mean | 59 \| 63 | 25 \| 26 | 25 \| 29 | 15 | 19 | 10 \| 10 | 15 \| 16 |
| Standard Deviation | 7 \| 6 | 4 \| 4 | 4 \| 4 | 5 | 4 | 3 \| 3 | 3 \| 3 |
| Minimum | 46 \| 50 | 19 \| 19 | 14 \| 19 | 6 | 11 | 3 \| 5 | 7 \| 12 |
| Maximum | 71 \| 73 | 32 \| 37 | 32 \| 41 | 31 | 30 | 18 \| 21 | 21 \| 25 |

**Table 2.** Poleward linear trends in Hadley cell terminus latitudes over 1981-2016. Units are °latitude decade$^{-1}$ with *p*-values in brackets (95% in bold).

|  | ERA-I | JRA55 | MERRA2 |
|---|---|---|---|
| ANNUAL NH GLOBAL | 0.2 (0.08) | 0.0 (0.71) | 0.0 (0.98) |
| ANNUAL NH WP | 0.4 (0.28) | 0.1 (0.87) | -0.4 (0.39) |
| ANNUAL NH EP | 0.2 (0.49) | -0.1 (0.60) | -0.6 (0.06) |
| ANNUAL NH NA | 0.0 (0.85) | 0.4 (**0.02**) | 0.3 (0.08) |
| ANNUAL SH GLOBAL | 0.2 (**0.02**) | 0.0 (0.72) | -0.1 (0.34) |
| ANNUAL SH SI | 0.1 (0.11) | 0.1 (0.20) | 0.3 (**0.03**) |
| ANNUAL SH SP | 0.6 (0.09) | 0.3 (0.48) | 0.1 (0.78) |
| SEASONAL NH GLOBAL | 0.3 (0.08) | 0.2 (0.31) | 0.3 (0.19) |
| SEASONAL NH WP | -0.0 (0.97) | -0.1 (0.86) | -0.3 (0.72) |
| SEASONAL NH EP | 0.3 (0.55) | -0.2 (0.61) | 0.2 (0.65) |
| SEASONAL NH NA | -0.0 (0.99) | 0.6 (0.13) | 0.7 (0.13) |
| SEASONAL SH GLOBAL | 0.3 (**0.00**) | 0.2 (**0.04**) | 0.3 (**0.02**) |
| SEASONAL SH SI | 0.4 (0.26) | 0.5 (0.16) | 0.8 (0.06) |
| SEASONAL SH SP | 0.8 (0.05) | 0.8 (0.07) | 0.6 (0.28) |

**Table 3.** As for table 2 but Hadley Circulation overturning absolute intensity with units $1 \times 10^9$ kg s$^{-1}$decade$^{-1}$.

|  | ERAI | JRA55 | MERRA2 |
|---|---|---|---|
| ANNUAL NH GLOBAL | 1.0 (0.20) | 5.5 (0.00) | 2.9 (**0.00**) |
| ANNUAL NH WP | 1.0 (0.65) | 7.6 (**0.00**) | -0.5 (0.89) |
| ANNUAL NH EP | -1.0 (0.77) | 10.8 (**0.00**) | 3.4 (0.28) |
| ANNUAL NH NA | 6.3 (**0.01**) | 8.2 (**0.00**) | 3.6 (0.08) |
| ANNUAL SH GLOBAL | 1.9 (**0.05**) | 1.0 (0.24) | 3.9 (**0.00**) |
| ANNUAL SH SI | 2.6 (0.11) | 8.1 (**0.00**) | 6.4 (**0.01**) |
| ANNUAL SH SP | 4.2 (0.18) | 7.9 (**0.01**) | 4.1 (0.37) |
| SEASONAL NH GLOBAL | 0.3 (0.66) | 1.6 (**0.00**) | -0.5 (0.38) |
| SEASONAL NH WP | 1.9 (0.54) | 4.3 (0.19) | 0.6 (0.87) |
| SEASONAL NH EP | -0.4 (0.92) | 8.0 (**0.04**) | 3.4 (0.36) |
| SEASONAL NH NA | 2.6 (0.06) | 2.7 (0.08) | -1.1 (0.51) |
| SEASONAL SH GLOBAL | 0.6 (0.52) | 0.7 (0.42) | 2.5 (**0.05**) |
| SEASONAL SH SI | -2.2 (0.46) | 7.0 (0.08) | 4.8 (0.18) |
| SEASONAL SH SP | 5.6 (0.21) | 5.5 (0.23) | 7.8 (0.10) |

LIST OF FIGURES

**Fig. 1.** The TC datasets and the ocean boundary definitions used in this analysis. (a) shows the ocean basin boundary definitions, note that the North Indian is excluded from this analysis for reasons explained in the text. (b) and (c) show all the TCs in the respective datasets from 1981 to 2016 by lifecycle point with genesis in blue, LMI in red and lysis in green. Note that the NHC+JTWC dataset reports a much earlier lysis point for TCs in the South Pacific.

**Fig. 2.** The zonally-asymmetric HC defined in the annual-mean divergent meridional overturning streamfunction at 500hPa and near-surface horizontal divergent winds in the ERA-Interim reanalysis for 1981-2016. Hadley cell termini are marked by the red curves.

**Fig. 3.** Quantile regressions for the meridional distribution of TC activity in both hemispheres: Northern (top: a, b, c) and Southern (bottom: d, e, f). The 95% confidence intervals are shown by the shaded regions. Trends from both dataset aggregations are shown: IBTrACS-WMO (lighter colors) and NHC+JTWC (darker colors). The maximum and minimum latitudes for each lifecycle point at the 20th, 40th, 60th and 80th percentiles are shown at the bottom of each subplot. The red horizontal lines shown the overall mean trends with the IBTrACS-WMO mean marked with triangles, and NHC+JTWC marked by squares.

**Fig. 4.** As in figure 3 but for the individual Northern Hemisphere ocean basins: the western North Pacific (top), the North Atlantic (middle) and the eastern North Pacific (bottom).

**Fig. 5.** As in figures 3 and 4 but for the individual Southern Hemisphere ocean basins: the South Pacific (top) and South Indian (bottom).

**Fig. 6.** Timeseries for HC extent diagnostic at each basin and hemisphere, computed in ERA- Interim, JRA55, and MERRA2.

**Fig. 7.** Regressions between seasonal-mean HC extent and seasonal-mean TC meridional distribution at different lifecycle stages. All timeseries had their linear trends removed. Pink shading shows regressions that exceed 95% confidence estimated using two-tailed *p* values from *t*-statistics. As there are no two independent observational records in the North Atlantic and eastern North Pacific, only the WMO data is shown for these basins.

**Fig. 8.** Timeseries for both hemisphere's seasonal-mean HC extent (blue) from the ERA-Interim Reanalysis and TC genesis LMI latitudes (red) from the NHC-JTWC observational record.

**Fig. 9.** Timeseries for the North Atlantic's seasonal-mean HC extent (blue) from the ERA-Interim Reanalysis and TC poleward [P(75)] lysis latitudes (red) from the NHC-JTWC observational record.

**Fig. 10.** As for figure 6, but for HC absolute intensity.

**Fig. 11.** Regressions between seasonal-mean HC intensity and seasonal-mean TC meridional distribution at different lifecycle stages. All timeseries had their linear trends removed. Pink shading shows regressions that exceed 95% confidence estimated using two-tailed *p* values from *t*-statistics. As there are no two independent observational records in the North Atlantic and eastern North Pacific, only the WMO data is shown for these basins.

**Fig. 12.** Timeseries for the eastern North Pacific's seasonal-mean HC intensity (blue) from the ERA- Interim Reanalysis and TC equatorward [P(10)] genesis latitudes (red) from the NHC-JTWC observational record.

**Fig. 13.** Regression coefficients for seasonal-mean (a) SST and (b) VWS (200 - 850 hPa wind vector difference) against the respective hemisphere's Hadley Circulation extent diagnostic 1981- 2016. The Northern Hemisphere in both plots is over JASO and the Southern Hemisphere JFM. All TC lifecycle points over the same period are overlain in green: genesis (a) and LMI (b)

**Fig. 14.** As for figure 13 but against eastern North Pacific local Hadley Circulation intensity diagnostic.

**Fig. A1**. Monthly mean meridional mass flux streamfunction derived with the divergent winds in ERA-Interim and the identified termini of the Hadley cells. (a) is the western Pacific in July 1980 and (b) the Atlantic in September 1989.

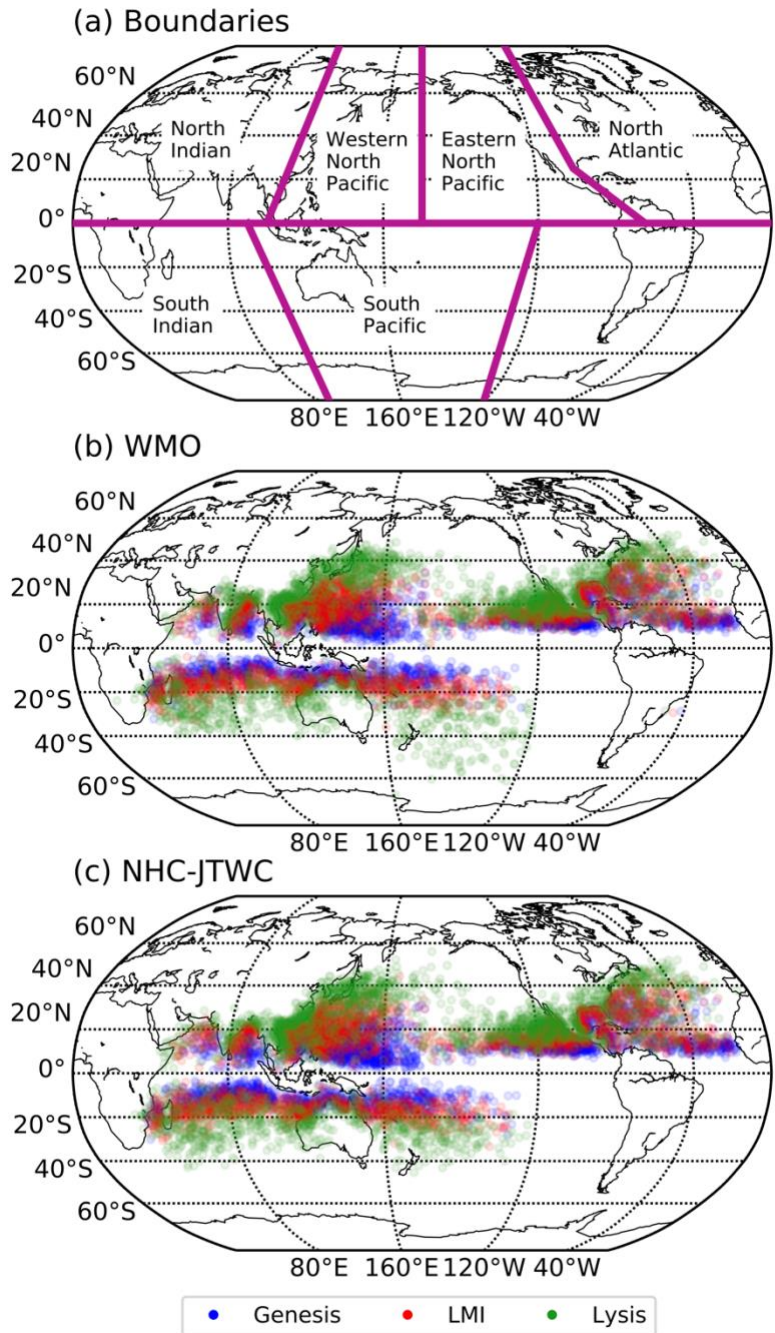

Figure 1: The TC datasets and the ocean boundary definitions used in this analysis. (a) shows the ocean basin boundary definitions, note that the North Indian is excluded from this analysis for reasons explained in the text. (b) and (c) show all the TCs in the respective datasets from 1981 to 2016 by lifecycle point with genesis in blue, LMI in red and lysis in green. Note that the NHC+JTWC dataset reports a much earlier lysis point for TCs in the South Pacific.

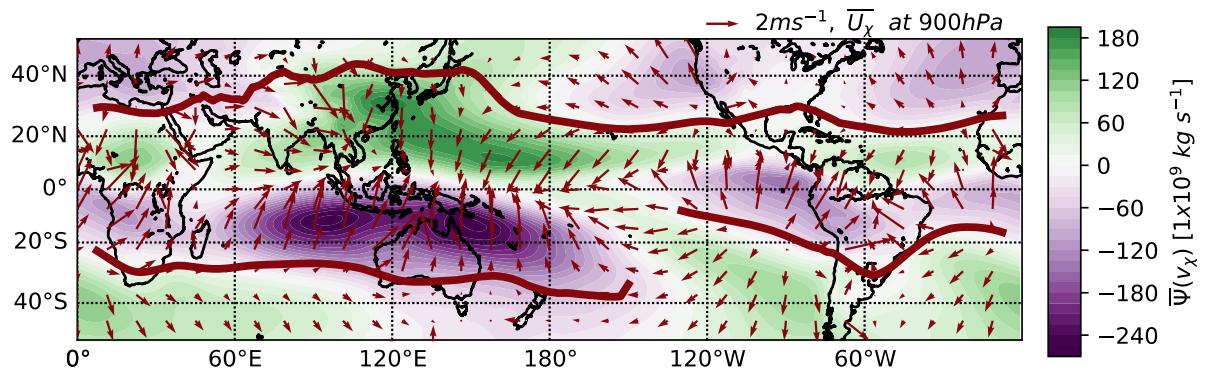

Figure 2: The zonally-asymmetric HC defined in the annual-mean divergent meridional overturning streamfunction at 500hPa and near-surface horizontal divergent winds in the ERA-Interim reanalysis for 1981-2016. Hadley cell termini are marked by the red curves.

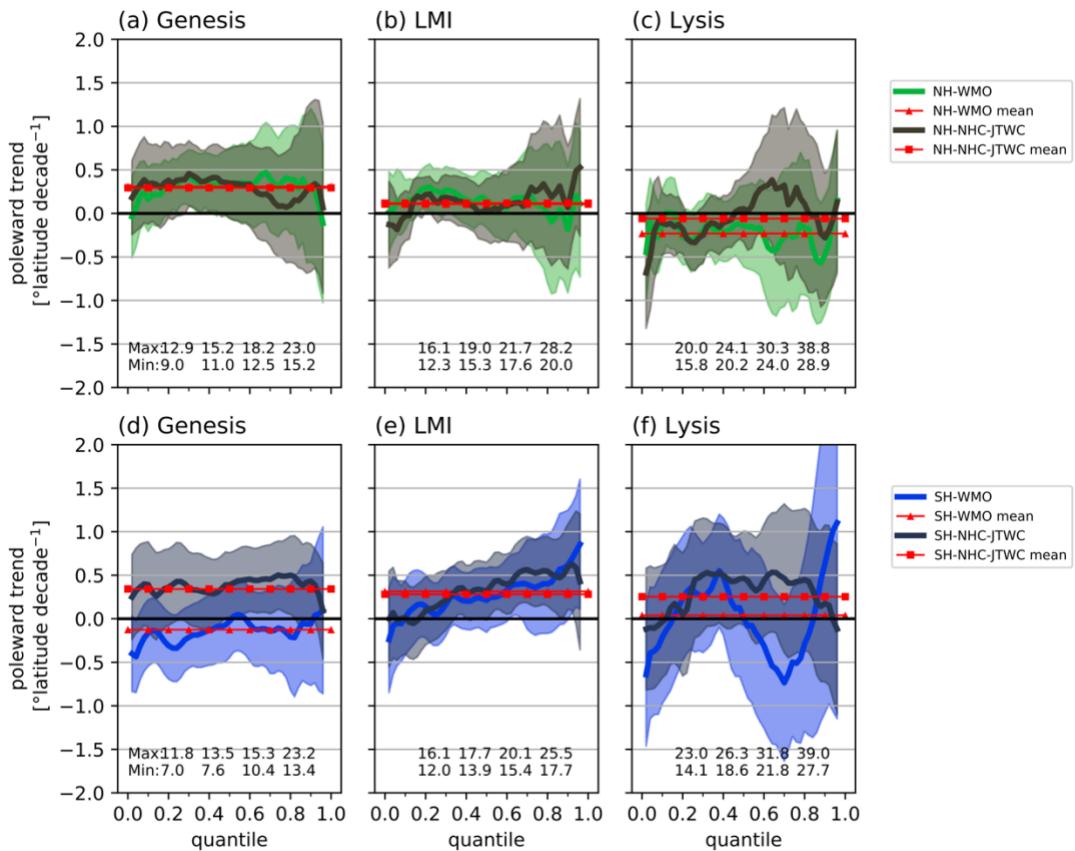

Figure 3: Quantile regressions for the meridional distribution of TC activity in both hemispheres: Northern (top: a, b, c) and Southern (bottom: d, e, f). The 95% confidence intervals are shown by the shaded regions. Trends from both dataset aggregations are shown: IBTrACS-WMO (lighter colors) and NHC+JTWC (darker colors). The maximum and minimum latitudes for each lifecycle point at the 20th, 40th, 60th and 80th percentiles are shown at the bottom of each subplot. The red horizontal lines shown the overall mean trends with the IBTrACS-WMO mean marked with triangles, and NHC+JTWC marked by squares.

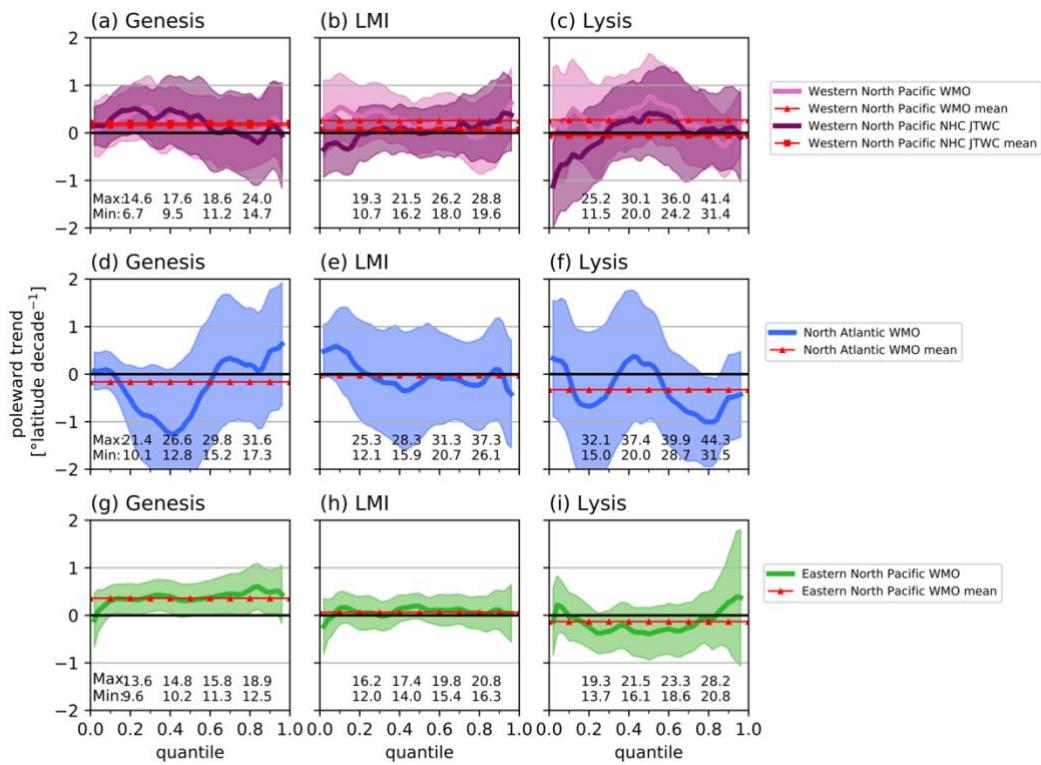

Figure 4: As in figure 3 but for the individual Northern Hemisphere ocean basins: the western North Pacific (top), the North Atlantic (middle) and the eastern North Pacific (bottom).

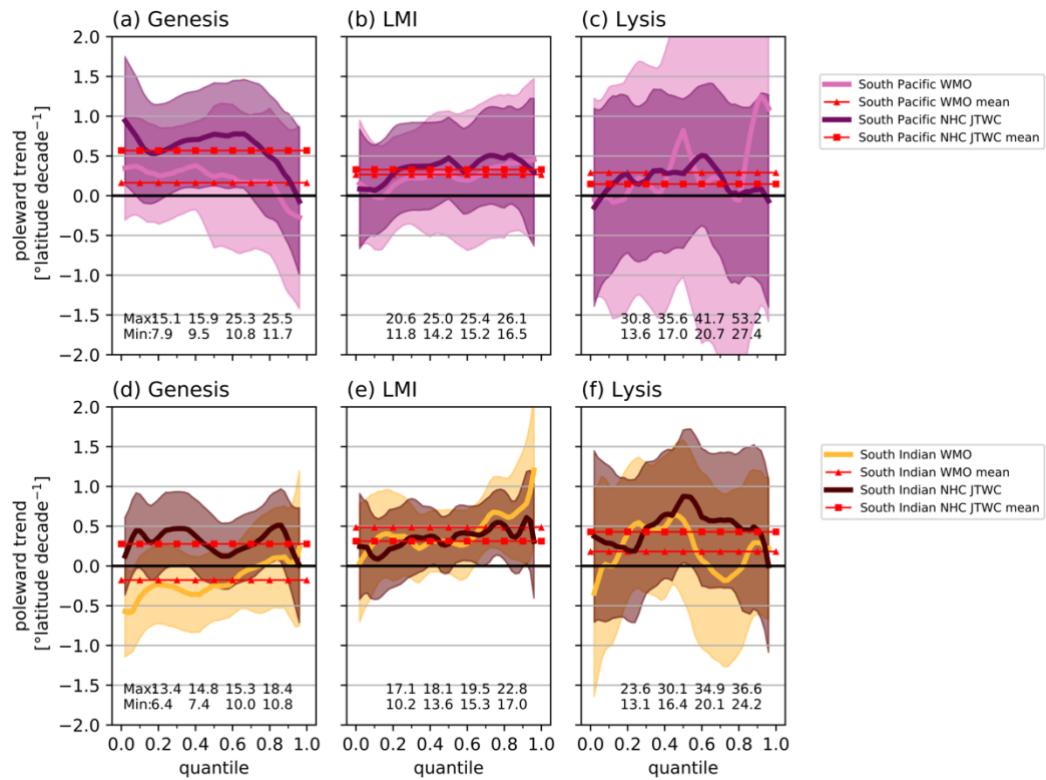

Figure 5: As in figures 3 and 4 but for the individual Southern Hemisphere ocean basins: the South Pacific (top) and South Indian (bottom).

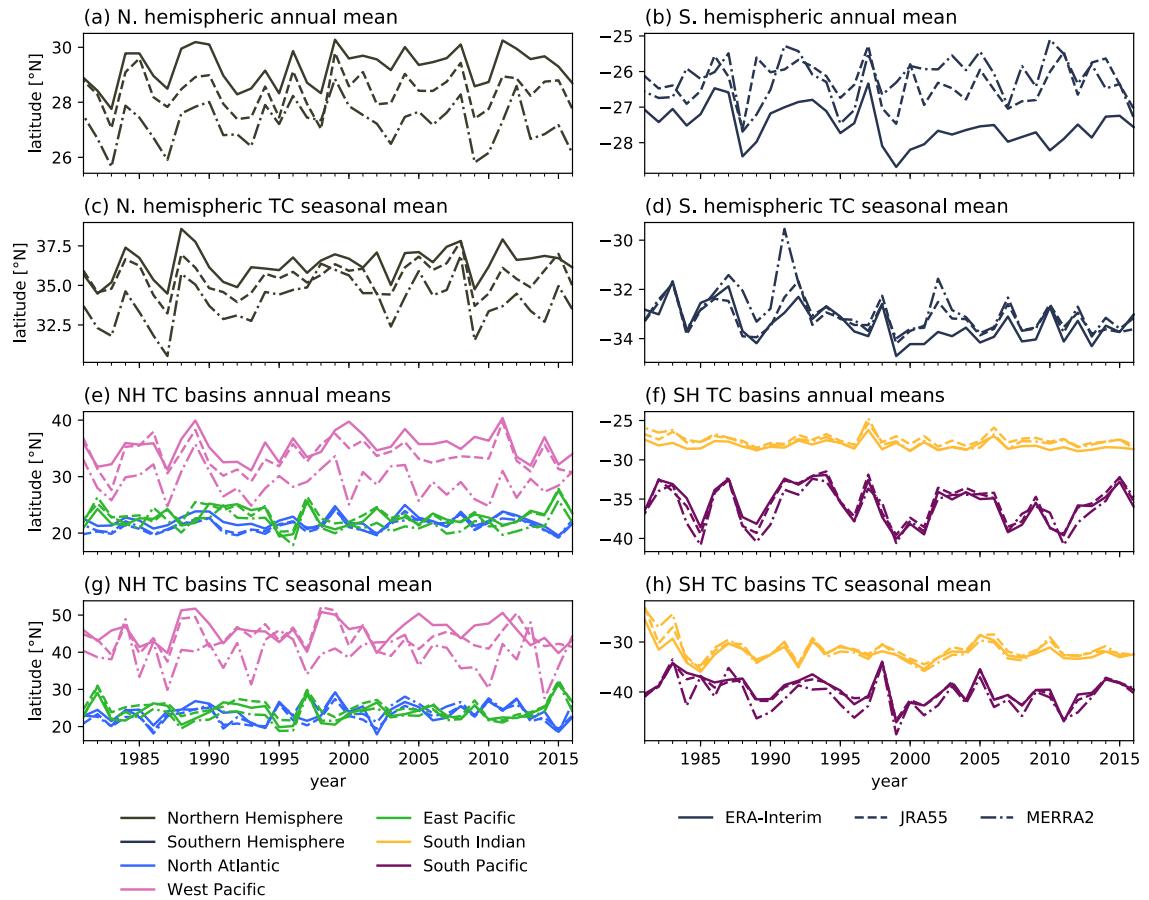

Figure 6: Timeseries for HC extent diagnostic at each basin and hemisphere, computed in ERA-Interim, JRA55, and MERRA2.

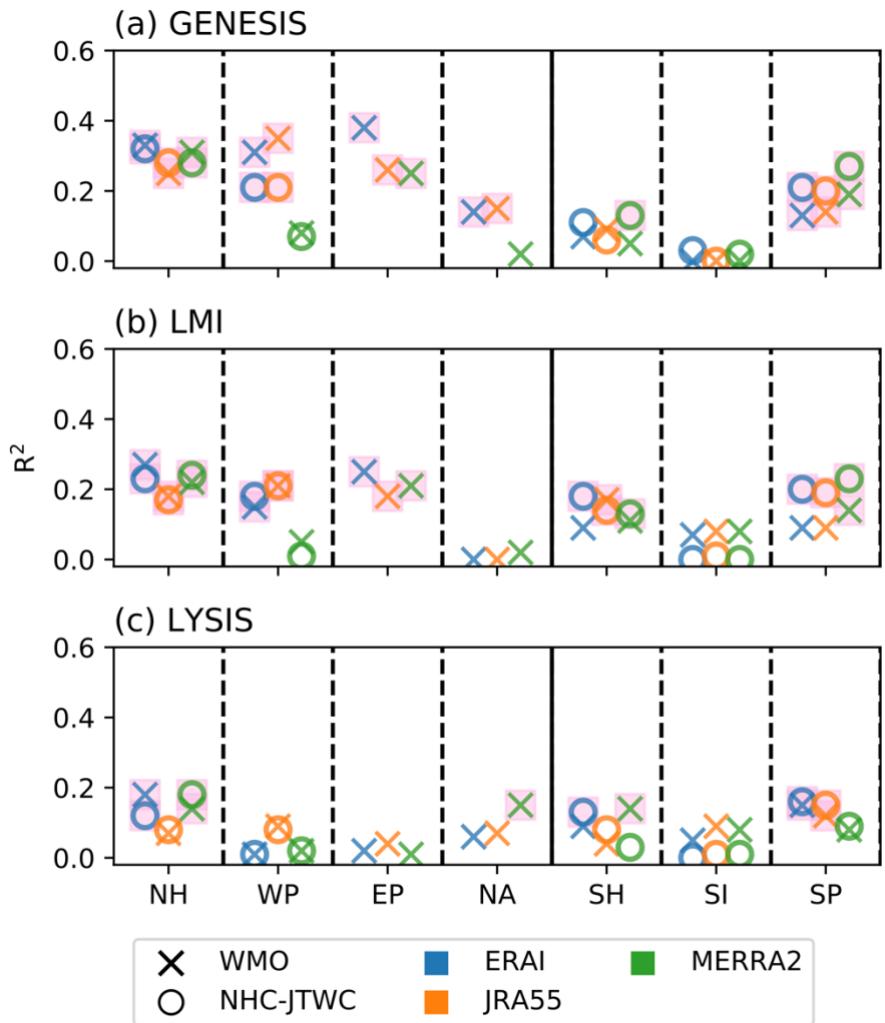

Figure 7: Regressions between seasonal-mean HC extent and seasonal-mean TC meridional distribution at different lifecycle stages. All timeseries had their linear trends removed. Pink shading shows regressions that exceed 95% confidence estimated using two-tailed *p* values from *t*-statistics. As there are no two independent observational records in the North Atlantic and eastern North Pacific, only the WMO data is shown for these basins.

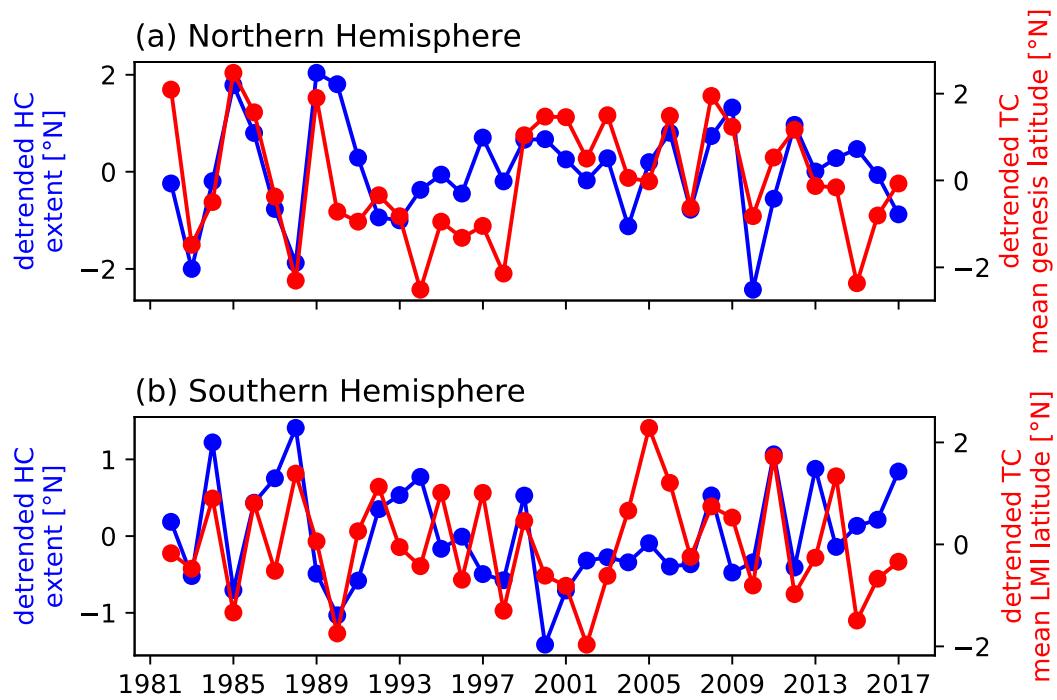

Figure 8: Timeseries for both hemisphere's seasonal-mean HC extent (blue) from the ERA-Interim Reanalysis and TC genesis LMI latitudes (red) from the NHC-JTWC observational record.

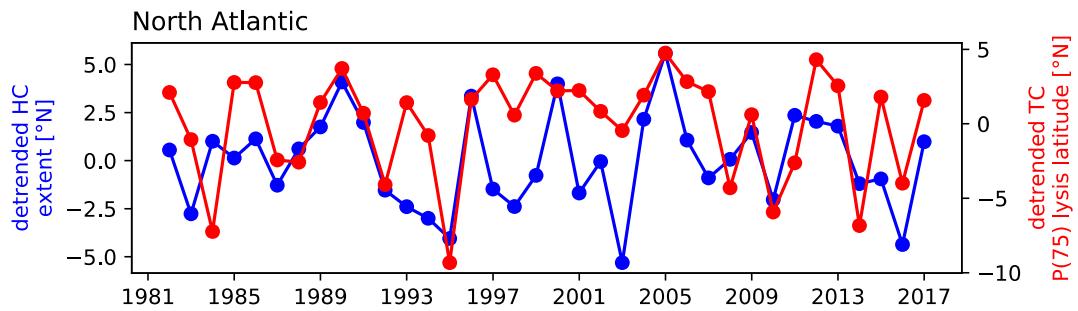

Figure 9: Timeseries for the North Atlantic's seasonal-mean HC extent (blue) from the ERA-Interim Reanalysis and TC poleward [P(75)] lysis latitudes (red) from the NHC-JTWC observational record.

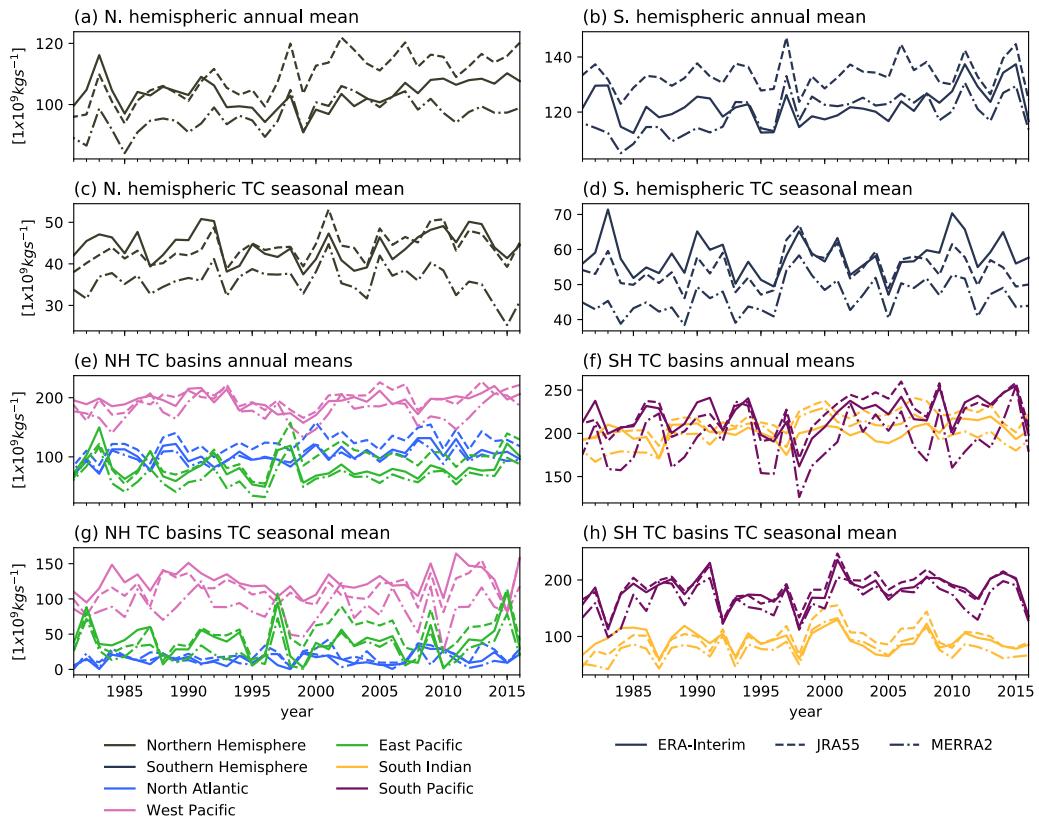

Figure 10: As for figure 6, but for HC absolute intensity.

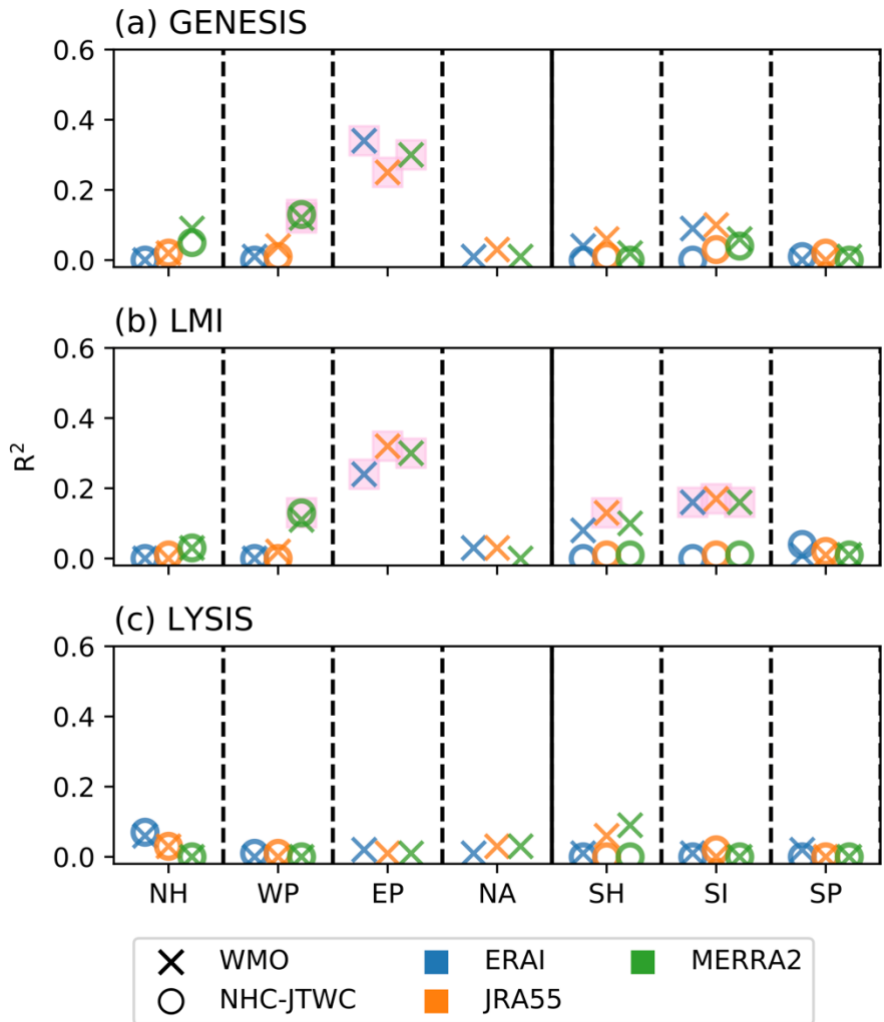

Figure 11: Regressions between seasonal-mean HC intensity and seasonal-mean TC meridional distribution at different lifecycle stages. All timeseries had their linear trends removed. Pink shading shows regressions that exceed 95% confidence estimated using two-tailed *p* values from *t*-statistics. As there are no two independent observational records in the North Atlantic and eastern North Pacific, only the WMO data is shown for these basins.

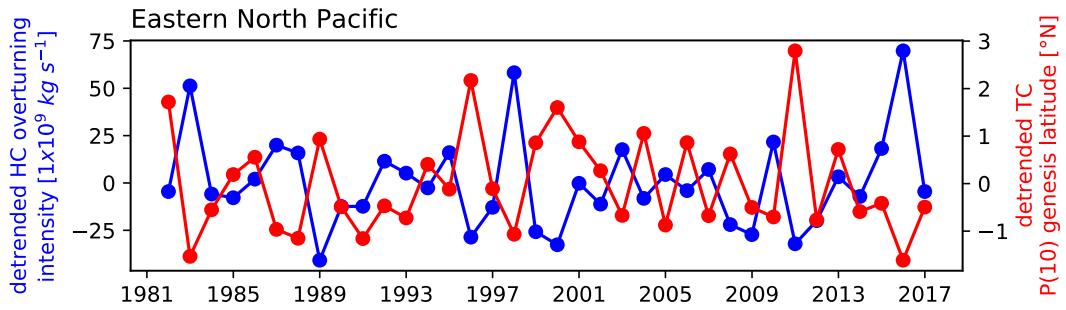

Figure 12: Timeseries for the eastern North Pacific's seasonal-mean HC intensity (blue) from the ERA- Interim Reanalysis and TC equatorward [P(10)] genesis latitudes (red) from the NHC-JTWC observational record.

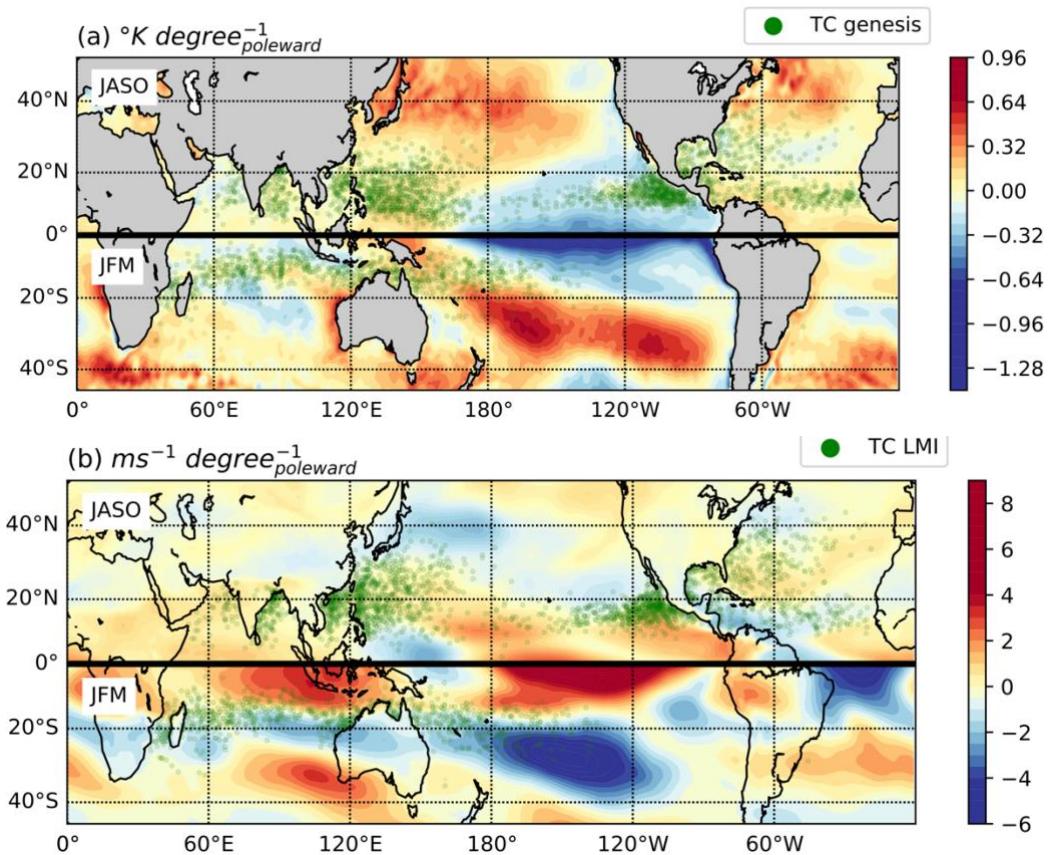

Figure 13: Regression coefficients for seasonal-mean (a) SST and (b) VWS (200 - 850 hPa wind vector difference) against the respective hemisphere's Hadley Circulation extent diagnostic 1981- 2016. The Northern Hemisphere in both plots is over JASO and the Southern Hemisphere JFM. All TC lifecycle points over the same period are overlain in green: genesis (a) and LMI (b).

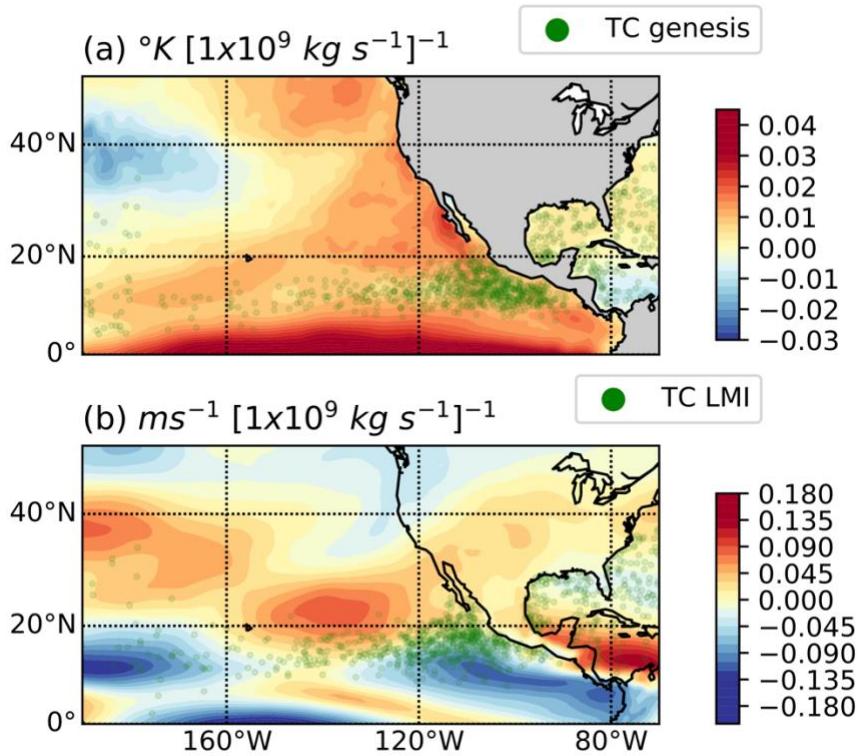

Figure 14: As for figure 13 but against eastern North Pacific local Hadley Circulation intensity diagnostic.

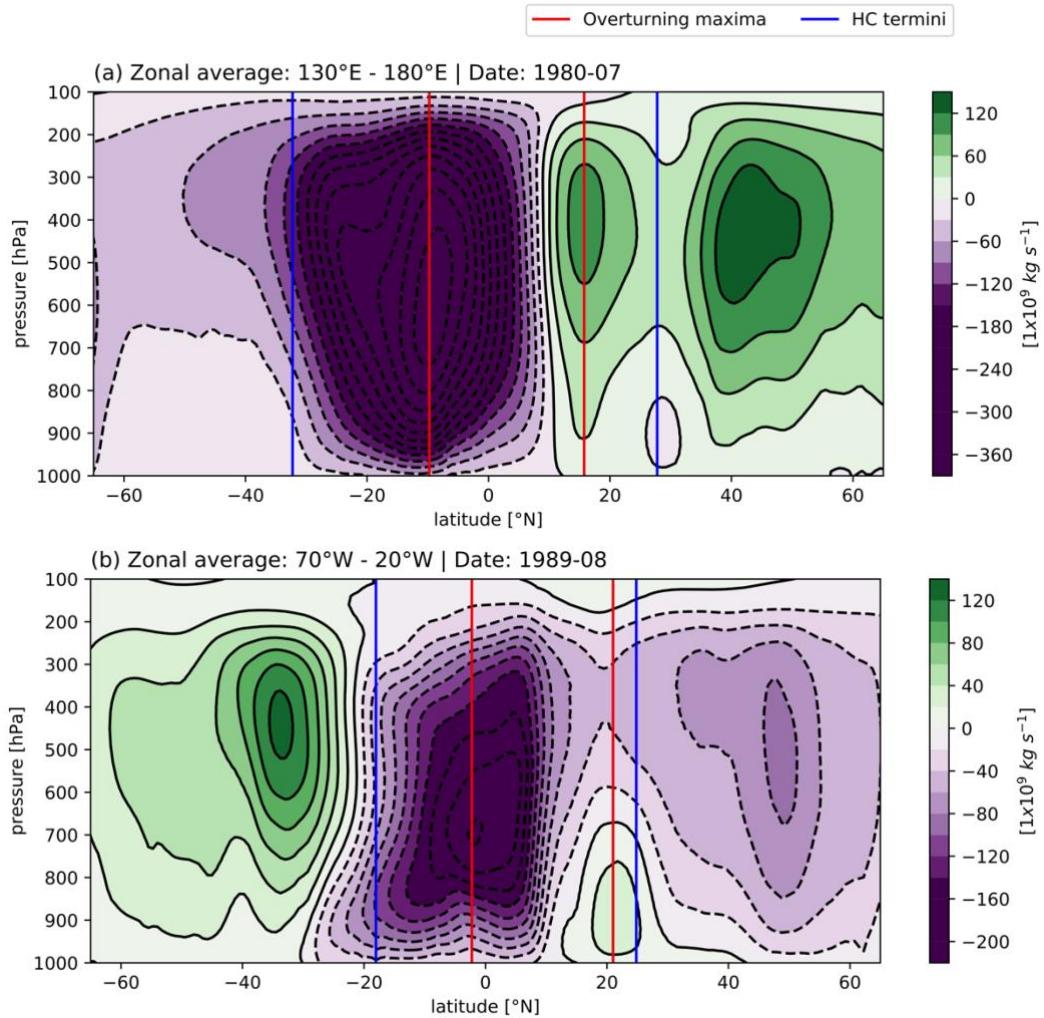

Figure A1: Monthly mean meridional mass flux streamfunction derived with the divergent winds in ERA-Interim and the identified termini of the Hadley cells. (a) is the western Pacific in July 1980 and (b) the Atlantic in September 1989.